# Line Defects in Molybdenum Disulfide Layers

*Andrey N. Enyashin,*[*a,b] *Maya Bar-Sadan*[c]*, Lothar Houben,*[d] *and Gotthard Seifert*[a]

[a] Physical Chemistry Department, Technical University Dresden, Bergstr. 66b, 01062 Dresden, Germany

[b] Institute of Solid State Chemistry UB RAS, Pervomayskaya Str. 91, 620990 Ekaterinburg, Russia

[c] Department of Chemistry, Ben Gurion University of the Negev, Be'er Sheva, Israel

[d] Peter Grünberg Institute, Ernst Ruska-Centre for Microscopy and Spectroscopy with Electrons, Research Centre Jülich, 52425 Jülich, Germany

enyashin@ihim.uran.ru;   barsadan@bgu.ac.il;   l.houben@fz-juelich.de;   gotthard.seifert@chemie.tu-dresden.de

ABSTRACT

Layered molecular materials and especially $MoS_2$ are already accepted as promising candidates for nanoelectronics. In contrast to the bulk material, the observed electron mobility in single-layer $MoS_2$ is unexpectedly low. Here we reveal the occurrence of intrinsic defects in $MoS_2$ layers, known as inversion domains, where the layer changes its direction through a line defect. The line defects are observed experimentally by atomic resolution TEM. The structures were modeled and the stability and electronic properties of the defects were calculated using quantum-mechanical calculations based on the Density-Functional Tight-Binding method. The results of these calculations indicate the occurrence of new states within the band gap of the semiconducting $MoS_2$. The most stable non-stoichiometric defect structures are observed experimentally, one of which contains metallic Mo-Mo bonds and another one bridging S atoms.

KEYWORDS    Grain boundaries; Monolayers; Density of states; Defects; High-Resolution Transmission Electron Microscopy; Density Functional Theory

Progress in chip design and advances of computational capabilities cannot go beyond the current strategy of integrated circuits based on classical IV or III-V semiconductors. The rise of graphene [1] – single-atomic sheet of layered carbon – suggested a possible way to overcome the limitations of widely spread photolithography by developing circuits based on layered compounds. Field effect transistors based on graphene suffer from its semi-metal nature, which results in small current on/off ratios despite the high intrinsic mobility and the presence of massless Dirac fermions in graphene. Therefore layers of d-metal dichalcogenides were proposed, among other candidates, as materials which can combine low dimensionality and suitable characteristics for cost-effective mounting of nanodevices [2].

Special attention is devoted to single layers of molybdenum disulphide, $MoS_2$, a well-established semiconducting compound with low toxicity and high thermal and chemical endurance. Bulk $MoS_2$ has an indirect band gap of 1.2 eV [3] and a rather high carrier mobility [4, 5]. In contrast to the bulk material, the observed electron mobility in single-layer $MoS_2$ is unexpectedly low [2, 6]. Point defects, dislocations or extended grain boundary defects may induce changes in the electronic structure that affect the transport of free charge carriers and may therefore be a primary source for the unexpectedly low mobility. The origin of such lattice defects in $MoS_2$ is related to the fabrication techniques. Single layers can be obtained from a bulk crystals by exfoliation [7, 8] or by mechanical cleavage [6]. This top-down approach produces a mixture of single layers and flakes with a few layers thickness. Bottom-up approaches by vapour deposition have been utilized to produce high quality layers with thickness control. Single-layered $MoS_2$ islands with sizes varying from dozens of atoms to dozens of nanometers [9] were produced by subsequent S and Mo atomic depositions and annealing on a metal surface. Large-area growth of $MoS_2$ thin films with scalable thickness and sizes up to several millimetres was achieved using a chemical vapour deposition method on $SiO_2$/Si substrates [10, 11] or using the thermolysis of (alkyl)ammonium thiomolybdates on a substrate [12]. Bottom-gated transistors constructed from such $MoS_2$ monolayers have often demonstrated lower carrier mobility than those built from mechanically cleaved monolayers. One possible reason for the reduced mobility is that the $MoS_2$ layers prepared by vapour deposition or thermolysis may consist of randomly oriented grains which contain numerous grain boundaries.

Despite their role in charge transport, the atomic structure and the formation mechanism of crystallographic faults in single layer dichalcogenides, in particular of non-stoichiometric defects with a deficiency or surplus of the metal or chalcogenide, have not been thoroughly investigated yet. Zou et al [13] predicted from first principles calculations structure and properties of edge dislocations and tilt grain boundaries formed by dislocation arrays that could emerge in the bottom-up growth and that include homo-elemental bonds. Here, we report on inversion domain boundaries in exfoliated MoS$_2$ and fullerenes which possibly occur as intrinsic defects in single layers that undergo stoichiometry changes in the sulfur shell. Atomic resolution images by aberration-corrected transmission electron microscopy provided evidence for these one-dimensional grain boundaries and a possible embedment of island-like domains enclosed by defect loops within a single MoS$_2$ layer. Quantum-mechanical calculations within the Density-Functional Tight-Binding method were used to analyze the stability and electronic properties of the defects. The results of these calculations indicate the occurrence of new states within the band gap of semiconducting MoS$_2$ as one of the possible factors affecting the electronic properties of MoS$_2$-based nanoelectronic devices.

Layers of dichalcogenides are characterized by a metal atom layer sandwiched between two atom layers of chalcogenides, with MoS$_2$ being a typical example [14]. The Mo atoms have a six-fold coordination environment, and in the most common phases the S atoms form a prismatic coordination around the Mo atom. In the most abundant polytypes of MoS$_2$ - 2H and 3R - two adjacent prismatic layers are stacked anti-parallel and parallel to each other, respectively (Fig. 1a). Stacking faults are known for the bulk material and were associated with dislocations along the [0001] axis, normal to the layers [15]. Moreover, the change of *inter-layer* orientation and the appearance of mixed stacking forms have been already reported for low-dimensional fullerenic-like structures of MoS$_2$ and WS$_2$ [16,17], where the edge-on configuration allowed direct imaging of the layers in a side-view.

In order to image *intra-layer* line defects, which occur within a single dichalcogenide layer, we have used atomic resolution high-resolution transmission electron microscopy (TEM). Both, a top view of a planar flake of MoS$_2$, consisting of a few layers, and a side-view of a curved structure (a MoS$_2$ onion-like structure) which allows an edge-on configuration of the layers, have been observed. Corresponding TEM images are shown in Fig. 1. The images were taken with an image-side aberration-corrected FEI Titan 80-300 microscope operated at 300 kV under negative spherical aberration imaging (NCSI) conditions [18,19]. These conditions produce high sensitivity for light elements and yield images that are a close representation of the projected potential for sufficiently thin samples, meaning bright dots can be correlated with the location of the atoms. The inversion of the MoS$_6$ prisms in one layer alters the stacking symmetry between two adjacent layers. Therefore, two layers, which were anti-parallel relative to each other on one side of a line defect, will acquire parallel stacking on the other side of the defect. In the case of anti-parallel stacking, the two layers will be imaged as the stable 2H stacking configuration, where hexagonal open rings are then clearly seen (see fig. 1a). The fingerprint of the parallel stacking (resembling the 3R lattice) will then be filled hexagons (fig. 1a). Experimentally, both empty and filled hexagons are imaged in close domains in the top view of commercially available 2H-$MoS_2$ powder (Sigma-Aldrich) exfoliated by sonication in iso-propanol (fig. 1b,c). In fig. 1b patches of filled hexagons and empty hexagons characteristic of 2H or 3R stacking are marked by white circles. Typically these domains form triangular islands within the layers. An enlarged view of the framed area (fig. 1c) contains an overlay of the different structure models of the 2H and the 3R. The white arrows in fig. 1b and the red circle in fig. 1c marks a grain boundary in a thin area of the sample, recognizable by the deviation from the perfect projection of hexagons observed in the adjacent domains with 2H and 3R stacking.

A quasi-spherical morphology of a nanoparticle allows observation of the cross-section of the layers, with direct evidence to the orientation of the $MoS_6$ prisms in the layers. Onion structures of $MoS_2$ cages were fabricated by solar ablation under conditions far from thermodynamical equilibrium, i.e. similar to those of the CVD technique [17]. Figs. 1d and 1e show atomic-scale images of the outer layers of a large $MoS_2$ fullerene. The prismatic coordination is then imaged by TEM as chevrons so that the layer direction as well as layer defects are easily identified [20,21]. An example for inversion domains is found in the atomic rows framed within fig. 1e, where a change of the chevron direction associated with a change of the $MoS_6$ prism orientation occurs within a layer, revealing the presence of a line defect. The top layer shows chevrons facing each other and the bottom layer shows chevrons facing away from each other.

We suggest that these frequently observed transitions from the parallel to the anti-parallel domains occurs via inversion domain boundaries, according to the following modeling and calculations. The calculations were performed using the density-functional based tight-binding (DFTB) method [22] with full geometry optimization and/or molecular dynamics (MD) simulations (NVT ensemble at T = 300K and 600K) within the Γ-point approximation and periodic boundary conditions, as implemented in the deMon software package [23]. In the DFTB approach, the single-particle Kohn-Sham eigenfunctions are expanded in a set of localized atom-centered basis functions, which are determined by self-consistent density-functional calculations on the isolated atoms employing a large set of Slater-type basis functions. The effective one-electron potential in the Kohn-Sham Hamiltonian is approximated as a superposition of atomic potentials, and only one- and two-center integrals are calculated to set up the Hamilton matrix. A minimal valence basis set has been used with 3s, 3p and 5s, 5p, 4d functions for S and Mo, respectively. It gives a correct description for the stability, electronic and mechanical properties of $MoS_2$ nanostructures in accordance with experimental data [20, 24-26].

For the stability estimations we used supercells of the monolayers of hexagonal $MoS_2$, which included 48 Mo and 96 S atoms, corresponding to 6×4 unit cells of an ideal lattice in a rectangular $a\sqrt{3}\times a$ representation. To exclude artificial interactions between the defects, we performed also test calculations on selected 4×4, 8×4 and 10×4 supercells. The results of these calculations showed that a 6×4 super cell is large enough to give converged defect formation energies. In order to preserve the overall stoichiometric composition, every unit cell of defect layers should contain two line defects, not necessarily of the same type. Therefore the calculated energies are related to the average of the combination of two defects. The energy of a structure was calculated per unit length, relative to the ideal prismatic $MoS_2$ layer.

As the starting point in finding the construction principles of inversion domains in $MoS_2$, we refer to the empirical Ostwald's step rule [27]. It claims that a compound tends to form a metastable state through the smallest loss of free energy before reaching its most stable state. We suggest that such a metastable state is the 1T allotrope of $MoS_2$, where Mo has an octahedral sulfur coordination [28]. In previous studies (see references in [28]) the 1T phase was found to be higher in formation energy with respect to the ideal prismatic $MoS_2$ layer by 19 meV/Å/atom. The defect nucleus with a formation energy of 23 meV/Å/atom which is shown in fig. 2a is the only combination of octahedral and prismatic $MoS_6$ units, which satisfies both the conditions of stoichiometry and coordination numbers of perfect 1T and 2H phases. The defect nucleus may be viewed as a line of prismatic coordination within a layer of octahedrally coordinated $MoS_2$.

The defect nucleus will re-arrange into two possible line defects, where Mo is coordinated to 4 or 8 S atoms in a planar coordination (fig. 2, structures I and II, respectively). Both coordination environments for Mo atom are atypical and not stable, with a calculated formation energy of 24.48 meV/Å/atom, therefore leading to further rearrangements.

Four-fold coordinated Mo atoms in the layers could rearrange into a line of Mo atoms with homonuclear Mo-Mo bonds – see fig. 2 (III). This rearrangement reduces the relative formation energy to about 4 meV/Å/atom. Other rearrangements are higher in energy and might even cause a rupture of the layer. MD simulations show that structure III with the Mo bridges is stable at 300 K and 600 K over 1 ns (see Movie in Supporting Information).

In turn, the sulfur excessive eight-fold coordination of Mo atoms such as in structure (II) is also unstable and may give rise to transformations into several related structures, including square-like defects and sulfur bridging atoms, all with rather low formation energies (2-3 meV/Å/atom). The most stable configuration consists of only –S– bridges (Fig. 2 IV) lines, which was found stable at 300 K over 1 ns and partial evaporation of –S– atoms is observed at 600 K.

Whereas various possible models of the line defects were constructed in the literature [13], we could relate our models to our experimental TEM results. The two stable configurations of these defects, the Mo bridges (III) and the S bridges (IV), can be distinguished from each other by looking at their side view projection. While the MoS$_2$ prismatic chevrons point away from each other in the Mo bridges (III), they point towards each other in the S bridges (IV) - see Fig. 2. Both configurations are present in the case of the fullerene-like particles: the two atomic layers framed in Fig. 1e show both chevrons facing each other (top layer) such as the –S- bridges and chevrons facing away from each other (bottom layer) such as the Mo bridges. The Mo-Mo bridges are detected in the top view images of the exfoliated layers in Fig. 1c within the red circle. A closer analysis of the TEM image contrast of this boundary on the basis of image calculations is given in Fig. 3. High-resolution TEM images were calculated for model structures containing the intra-layer defects in order to confirm the correspondence between our atomistic models and the experimental data considering the effect of sandwiching matrix layers. A forward-calculation of TEM images is required since the structure is not periodic along the viewing direction, unlike the case of the fullerene layer in side-view. The simulations are based on multislice calculations [29]. Super-cells comprising a single layer that contains the line defect and sandwiches of this layer with enclosing MoS$_2$ layers were used for the multislice iteration. Up to three matrix layers in 2H stacking symmetry were added, resulting in a maximum of four MoS$_2$ layers. Images were then calculated for the optimum negative phase contrast settings (NCSI) in an aberration-corrected TEM [18,19] operated at 300 kV, corresponding to experimental conditions. Fig. 3 shows the simulated images for the case of line defects containing Mo-Mo bonds and a line defect containing -S- bridges, models III and IV according to Figure 2, as well as the experimental image reproduced and magnified from Fig. 1c. For both cases, the defect with Mo-Mo bonds in the top row of Fig. 3 and the defect with the –S- bridges in the bottom row, the coordination of the defect is directly visible from HRTEM images only for the single layer case. As soon as there are matrix layers on top or at the bottom of the defect layer, the atomic coordination cannot be inferred directly from the image anymore. For a large number of matrix layers the defect becomes almost invisible since the periodic matrix contrast dominates. In thick layers one will therefore only observe domains with 3R or 2H stacking but no clear image of the boundary. However, the linear defect shows in two parallel chains of atoms in projection (see rectangular markers) along the traces of two neighboring (1-100) planes as long as there are only few matrix layers. Along these planes, the projection of hexagons that are typical for the bulk $MoS_2$ lattice is distorted since the atoms are not aligned to the columns of the matrix structure in viewing direction in the strain field around the defect. The experimental image to the right of Fig. 3 shows a very similar distortion in two neighboring planes (see rectangular marker) where there is no hexagon projection. The dense spacing of bright dots in the two neighboring atom chains in projection points to a higher density of Mo atoms rather than to the lower density in the case of the defect with -S- bridges. The experimental contrast pattern matches best for the case of the defect with Mo-Mo bonds with one or two enclosing matrix layers. While an exact match between the simulated pattern and the experimental pattern is not expected because it requires a precise knowledge of the strain relaxation of the defect in experiment, we find that the model of the Mo-Mo bond defect is consistent with the experimental data.

The appearance of defects may significantly influence the electronic properties and particularly the electronic transport properties. For the study of the influence of grain boundaries on the electronic properties the electronic structure of the ideal $MoS_2$ monolayer and the structures containing Mo-Mo or -S- line defects was studied. We examined supercells containing the loops of line defects: a triangular island of $MoS_2$ bound to the surrounding $MoS_2$ by either only S-bridges (IV) or only Mo-bridges (III). The stoichiometry of these cells is $Mo_{303}S_{586}$ and $Mo_{293}S_{606}$, respectively and their optimized geometries are shown in Fig. 4. The geometry optimization of both models shows an induced strain within the domain structures, since the unit cells are contracted by 4% at this size as compared to the ideal $MoS_2$. Nevertheless, both structures keep their integrity during the MD simulations at 300K and can therefore be recognized as stable.

The calculated electronic density-of-states of a perfect single prismatic $MoS_2$ layer (Fig. 5a) shows three main features, the Mo$4d$-S$3p$ valence band (ranging approximately from -7 eV to -4 eV), the Mo$4d_z^2$ band around -1.5 eV and the conduction band states above ~1 eV. This result agrees well with former experimental and theoretical investigations [7,14,28]. In earlier studies of the $MoS_2$ monolayer or nanotubes only zero-dimensional defects (vacancies, substitutional dopants, square-like defects) were considered [30,31]. These defects mainly create localized states within the band gap of pristine $MoS_2$ or cause a shift of the Fermi level into the Mo$4d_z^2$ band.

The densities of states (DOS) for the optimized cells of the structures containing the line defects are shown in Fig. 5b and 5c. In general, the DOS profiles of these hybrid systems are quite similar to that of the ideal $MoS_2$ layer, but there is a nonzero density of states in the region of the band gap of pristine $MoS_2$. The partial DOSs reveals that the main contribution to these states stems from $Mo4d$-states of both the island atoms and the surrounding atoms, around the island. Moreover, a detailed view of the wave functions for the crystal orbitals reveals a large contribution to the Fermi level from the orbitals of the Mo atoms placed near the edges of both inverse domains (Fig. 5).

In the case of the domain structure, which contains Mo-Mo line defects, the contribution of $Mo4d$ orbitals from the interlinking atoms to the HOCO and the LUCO is observed, as well as into the $Mo4d_z^2$-band (Fig. 5b and pdf-file in Supporting Information). Furthermore, the $Mo4d$ states that originate mainly from the interlinking atoms form a new band at 2.5 eV below the Fermi level. The new band accounts for the formation of stable covalent Mo-Mo bonds. In contrast, in the structure that contains the two-fold -S- bridges, the line defects produce $S3p$-states that do not contribute to the DOS near the Fermi level (Fig. 5c). They can be found in the valence band region at 3 eV below Fermi energy. Thus, the states near the Fermi level arise exclusively from the edge states like in free standing triangular $MoS_2$ nanoplateletes [32,33].

In conclusion, we were able to reveal by HRTEM and to elucidate by quantum-mechanical calculations the line defects within a molecular sheet of $MoS_2$. The line defects separate patches or islands, where the layer direction is opposite to its surrounding. Using DFTB molecular dynamics simulations, we identified two atomic structures as most probable, structures with two antiparallel $MoS_2$ parts interlinked by either Mo-Mo bonds or by -S- bridges. We could compare and identify experimentally by HRTEM the features of both defect types.

The latter allows us to assume the formation mechanism of these particular defect types as those arising from a defective 1T-phase as possible ancestor. Certainly, there may be a wealth of possible defects with different growth mechanisms depending on particular experimental setup. Recently, other types of line defects have been observed in CVD grown $MoS_2$ layers [34,35], which represent the sets of squares and octagons or pentagons and heptagons among hexagonal phase of $MoS_2$ ("4-4-8" and "5-7" defects). While their origin remains unexplained, we can speculate that they may be as well the products of complex multi-stage reconstruction arising from 1T-layer with other types of defects. Yet, their formation also could be related with the multi-stage growth of a $MoS_2$ layer due to portioned delivery of Mo-S vapor to a growing seed. In this case the search of possible formation mechanism should imply not only an analysis of the defects' stability, but also the stability of different MoS$_2$ edges and energies of their saturation. In any case, the models of defects observed in [34,35] are not consistent with HRTEM images observed in our study due to different orientation of MoS$_2$ grains at grain boundary and due to the mismatch in the widths of line defects.

The domain boundaries might explain the differences of the electronic properties of single layers of MoS$_2$ relative to the pristine prismatic MoS$_2$ layer. All defects considered in our work form localized trap states near the Fermi level and in the band gap region. The electronic and transport properties of layers containing such defects stem mainly from the edge states of the two opposite domains, in a similar way as described in the literature for triangular MoS$_2$ nanoplatelets and nanostripes with metallic-like edges. Depending on factors such as the concentration of line defects, their distribution and mutual orientation such states may influence the overall charge carrier mobility of the material, which can explain the variations of the charge carrier mobility in the single layer MoS$_2$ devices. Yet, the detailed influence of grain boundaries on the transport properties of 2D MoS$_2$ layers requires a special attention.

**Acknowledgment.** The support of the ERC project INTIF 226639 is gratefully acknowledged. This research was supported by a grant from the GIF, the German−Israeli Foundation for Scientific Research and Development. A.E. thanks RFBR, grant 11-03-00156-a.

**Supporting Information Available.** Movie of molecular dynamics simulation at 300 K of MoS$_2$ supercell containing intrinsic one-dimensional defects: line of Mo-Mo dimers (on the left) and line of S bridging atoms (on the right) (SI_movie.avi); Γ-point isosurfaces of some crystal orbitals for the models of grain boundaries in MoS$_2$ formed by inversion domains connected by -S- line defects or via Mo-Mo line defects (SI_orbitals.pdf).

TOC

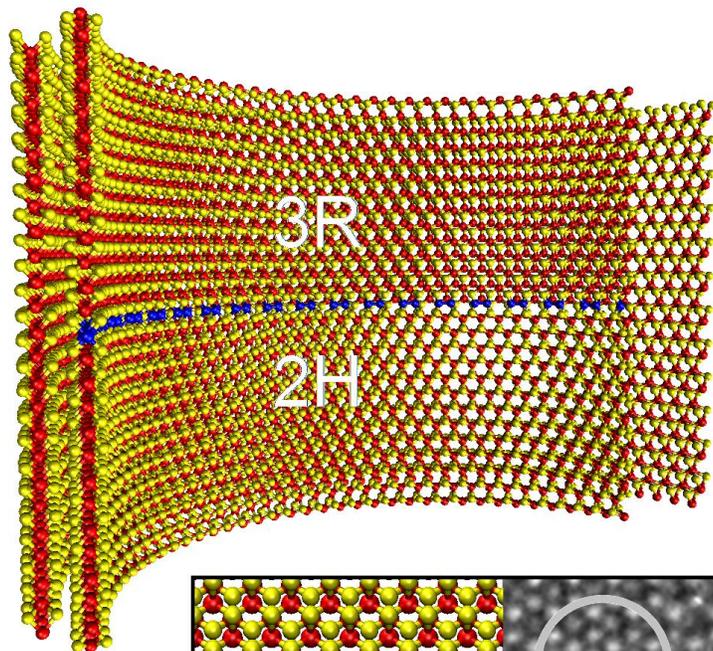

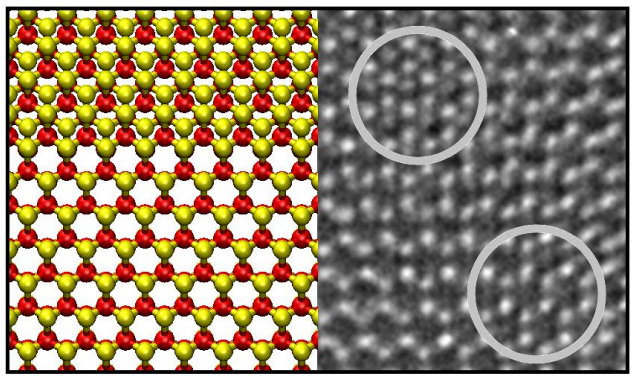

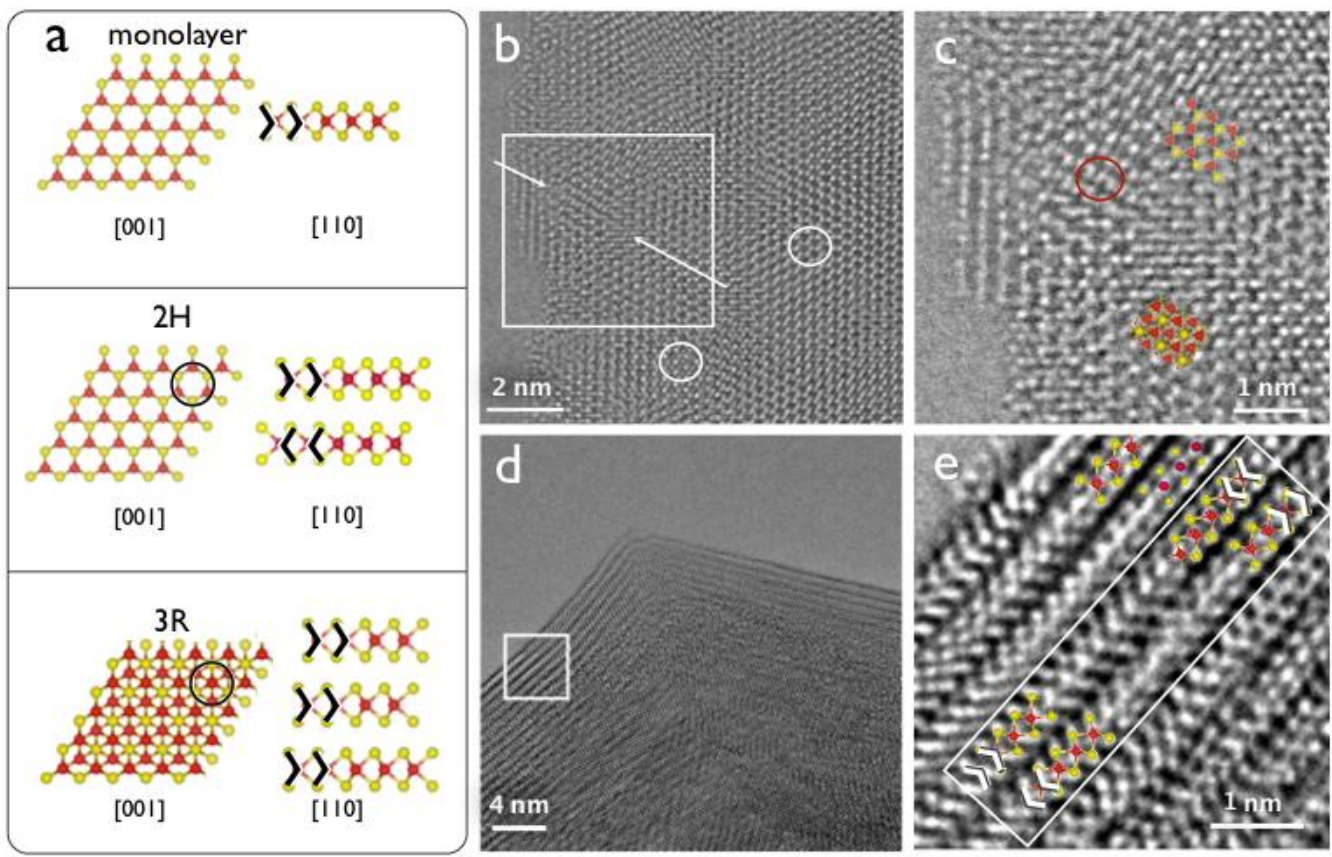

**Figure 1.** Structure models of the MoS$_2$ phases and high-resolution TEM image of line defects in MoS$_2$. (a) Top and side view of a monolayer, the 2H phase and the 3R phase of MoS$_2$. Mo in red and S in yellow. Angular markers (> and <) mark characteristic chevron-like S-Mo-S bonding in side view. In top view ([001] projection) the 2H produce open hexagons while the 3R phase shows filled hexagons, marked by the overlaid circles. (b) Top view of a stack of a few MoS$_2$ layers exfoliated from bulk material. Alternating domains of filled hexagons (like in the 3R stacking) and empty hexagons (like in 2H stacking) are marked by the circles. Arrows point to a line defect in a thin area. (c) A magnified view of (b). Structure models are overlaid on the TEM image, and a red circle marks the area of the defect where the deviation from a perfect crystal is seen as elongated features. (d) Edge view of a solar ablated MoS$_2$ fullerene. (e) A magnified part of the white frame presented in (d) reveals atypical stacking order. In particular the two layers inside the rectangular marker in (e) exhibit a change in the orientation of the chevron-like S-Mo-S pattern within one layer, i.e. an inversion of the orientation of MoS$_6$ prisms. The top layer shows the S bridges, where chevrons face each other and the atomic model is in accordance (overlaid). The bottom layer shows chevrons facing away from each other as in the Mo bridges model. [figs. 1d and 1e adapted from Ref. 17]

.

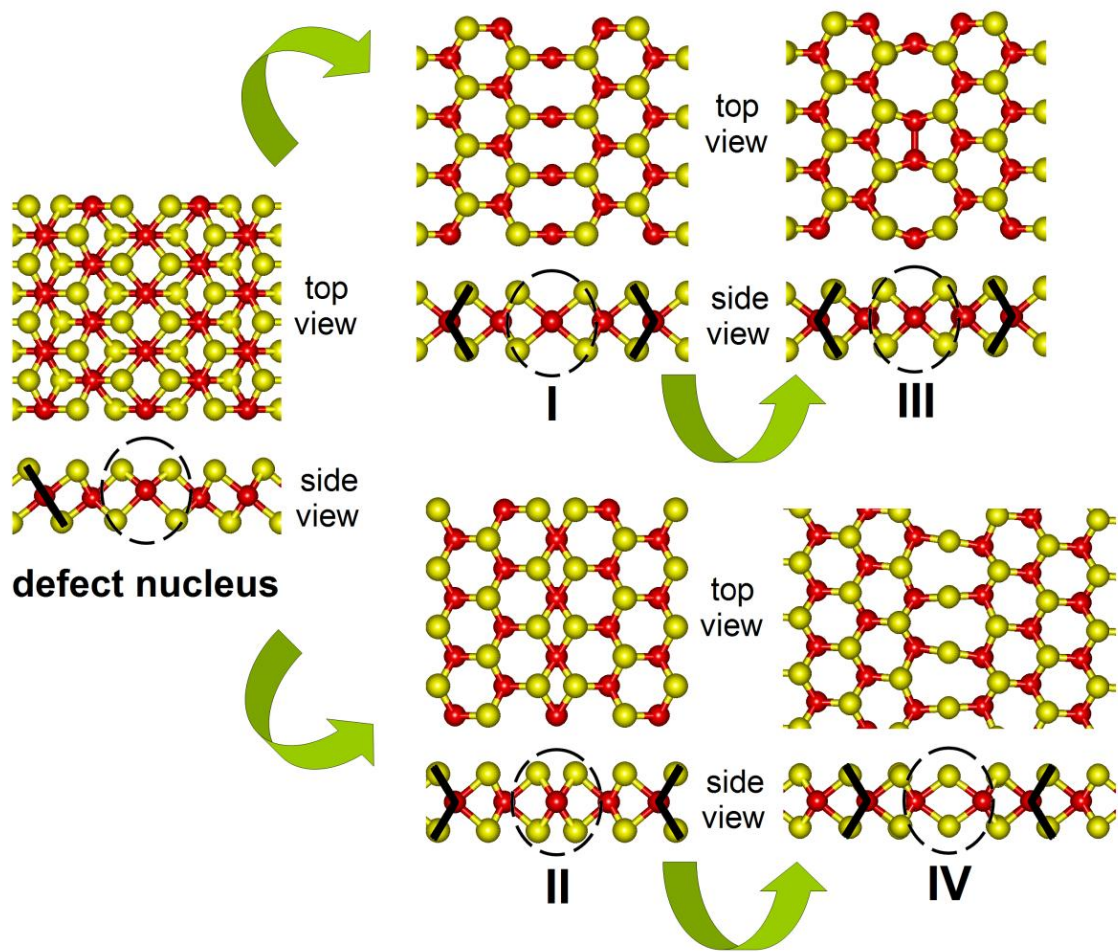

**Figure 2.** Models of MoS$_2$ layer fragments containing line defects and optimized using DFTB method. The defect nucleus of 1T-layer composed of MoS$_6$ octahedra interlinked with line prismatic MoS$_6$ prisms (on the left). The defect nucleus rearranges to a line defect of four-fold coordinated Mo atoms (I) and then to a line defect containing Mo-Mo metallic bonds. Alternatively, the defect nucleus may rearrange to a line defect of eight-fold coordinated Mo atoms (II) and then into a line defect composed of S bridging atoms (IV). All the line defects are marked by the dashed circle on the side view, and black chevrons are overlaid on the layers to guide the eye to the layer direction.

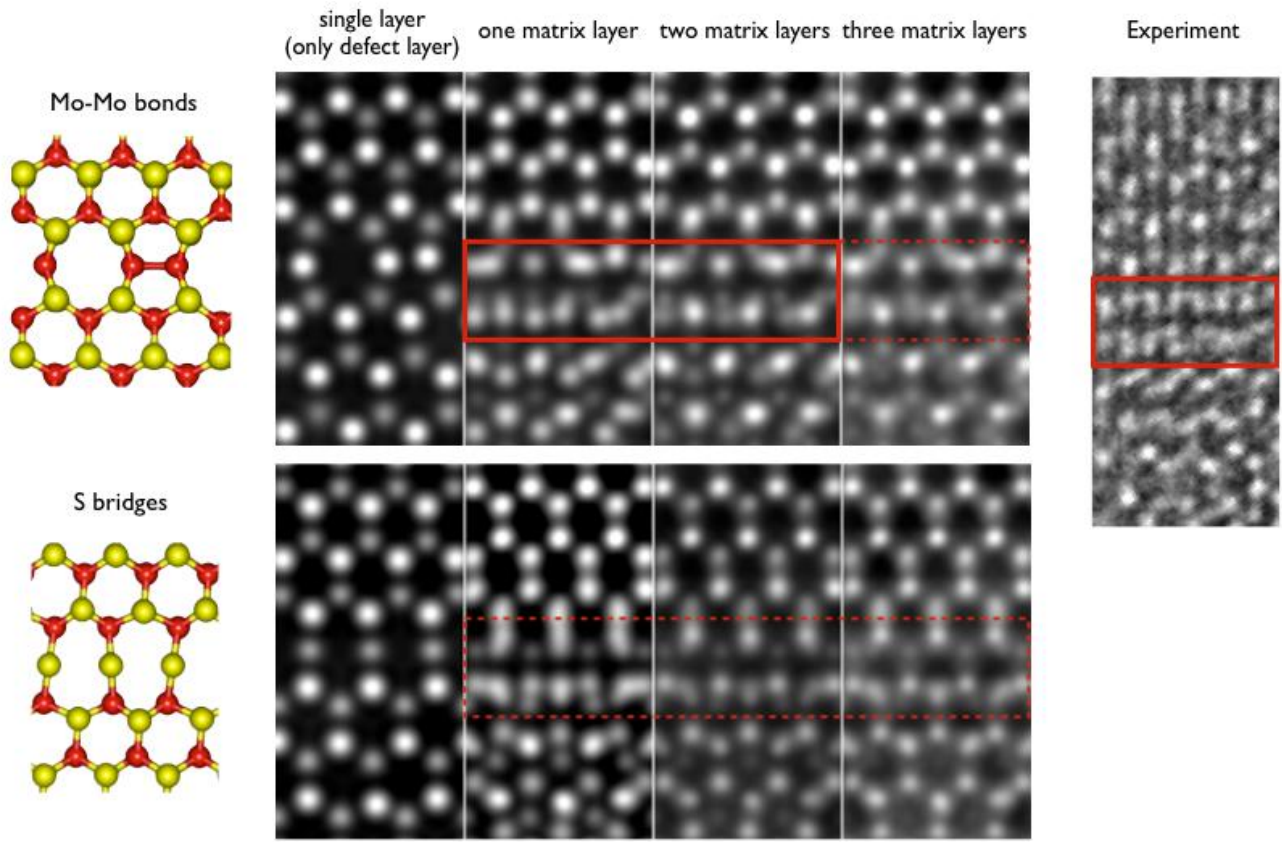

**Figure 3**. Calculated and experimental high-resolution TEM images of line defects in MoS$_2$. The left part of the figure shows simulated images for the Mo-Mo bond defect and the –S- bridges defect according to the models III and IV in Fig. 2. Images are shown for a single layer with a defect and for the defect layer sandwiched with addition matrix layers of undistorted 2H MoS$_2$. The right-hand side shows a magnified view of an experimental image taken from Fig. 1c. The rectangular markers expose two neighboring (1-100) planes where the defect causes the deviation from a hexagon projection. The experimental data shows a consistent match with the simulated images for the defect with Mo-Mo bond when taking one or two matrix layers into account.

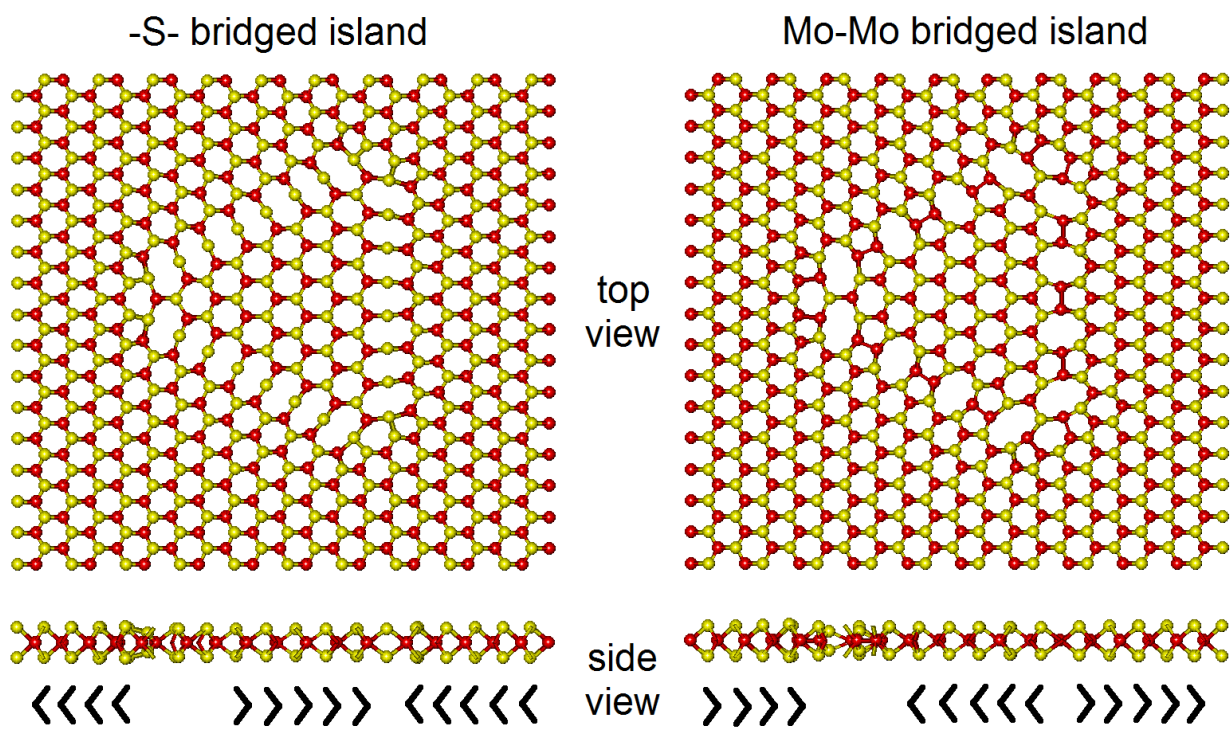

**Figure 4.** Grain boundaries in MoS$_2$ formed by inversion domains connected by -S- line defects (left) or via Mo-Mo line defects (right). The models were obtained using DFTB method. Mo in red, sulphur in yellow. The side cross-sections show both ball-and-stick and schematical chevron-like representations, which coincide with the characteristic patterns of inversion domains observed experimentally (see fig. 1).

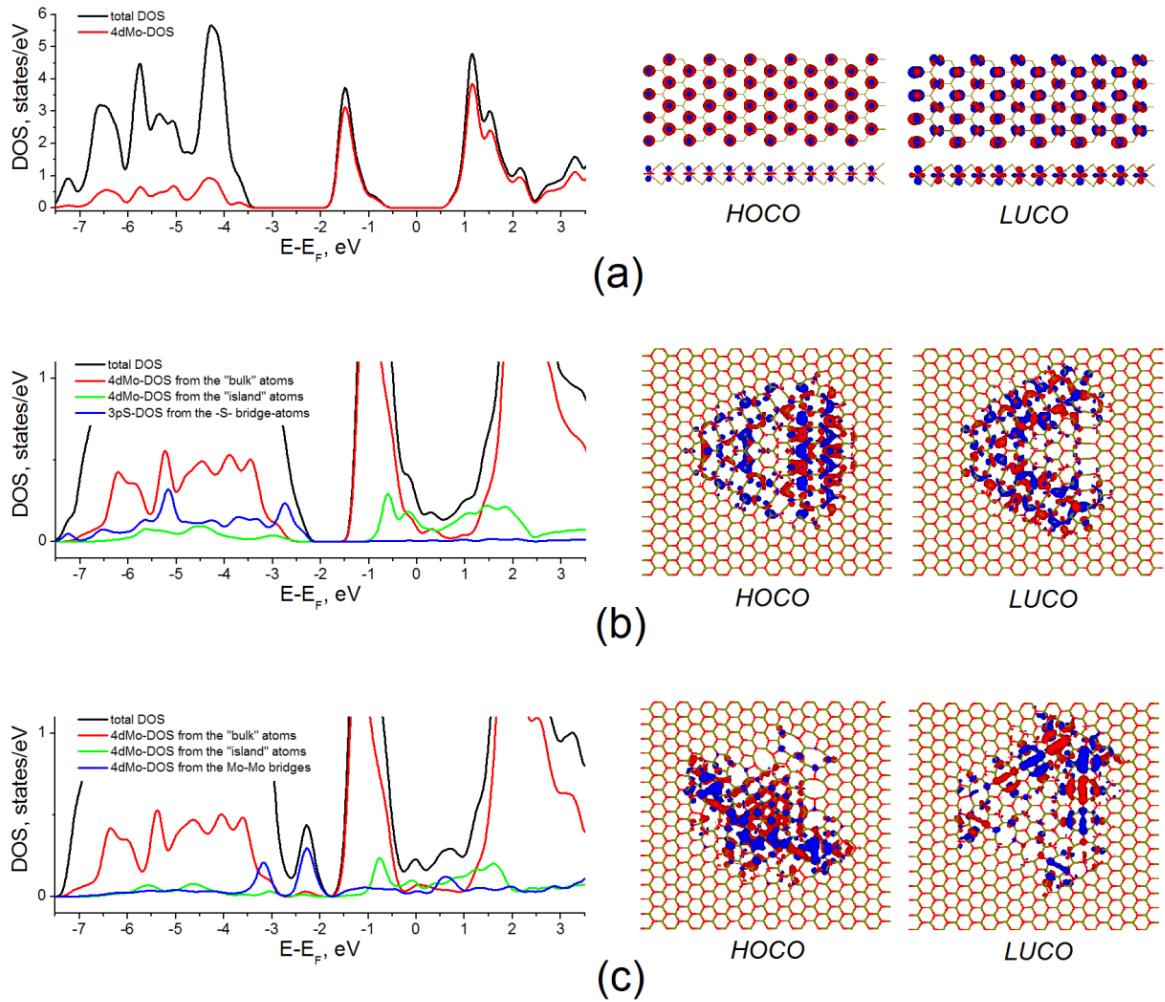

**Figure 5.** Total and partial densities of states (DOS) calculated using DFTB method for ideal monolayer of 2H-MoS$_2$ (a) and 2H-MoS$_2$ derived monolayers with the embedded islands interlinked either via -S- bridges (b) or via Mo-Mo bonds (c). Γ-point isosurfaces of highest occupied and lowest unoccupied crystal orbitals (HOCO and LUCO) are also depicted. The ball-and-stick models corresponding to (a-c) are depicted on figs. 1a and 3, respectively.